\documentclass[12pt]{article}
\usepackage{amsmath}
\usepackage{graphicx}
\usepackage{enumerate}
\usepackage{natbib}
\usepackage{url} 

\newcommand{\blind}{0}

\usepackage{amssymb,amsfonts,mathrsfs,amsmath,amsthm,mathtools,bm,dsfont, thmtools,bbm}\allowdisplaybreaks
\usepackage[dvipsnames]{xcolor}
\usepackage{array, graphicx, graphics,float,multirow,tabularx,tikz,pgfplots, longtable,threeparttable}
\usepackage{setspace}
\usepackage[colorlinks,citecolor=blue,urlcolor=blue, runcolor=blue, filecolor=cyan]{hyperref}
\usepackage{footnotebackref}
\usepackage{bibentry, natbib} 
\usepackage{paralist}  
\usepackage{appendix} 
\usepackage{mathrsfs}
\usepackage{enumitem}
\usepackage{authblk}
\usepackage{caption}
\usepackage{titlesec}
\usepackage{textcase,relsize}
\usepackage{datetime}
\usepackage{booktabs}

\declaretheoremstyle[notefont=\bfseries,notebraces={}{},%
    headpunct={},postheadspace=1em]{mystyle}
\declaretheorem[style=mystyle,numbered=no,name=Assumption]{asmp-hand}
\declaretheorem[style=mystyle,numbered=no,name=Condition]{cond-hand}
\declaretheorem[style=mystyle,numbered=no,name=Example]{exmp-hand}

\usepackage{titlesec}

\titleformat{\paragraph}
{\normalfont\normalsize\bfseries}{\theparagraph}{1em}{}
\titlespacing*{\paragraph}
{0pt}{3.25ex plus 1ex minus .2ex}{1.5ex plus .2ex}

			\def \bfA {\mathbf{A}}
	\def \calB {\mathcal{B}}		
	\def \calC {\mathcal{C}}		
			
\def \bbE {\mathbb{E}}			\def \bfE {\mathbf{E}}
			\def \bfF {\mathbf{F}}
	\def \calG {\mathcal{G}}		
			\def \bfH {\mathbf{H}}
	\def \calI {\mathcal{I}}		\def \bfI {\mathbf{I}}

			\def \bfM {\mathbf{M}}
	\def \calN {\mathcal{N}}		
			
			\def \bfP {\mathbf{P}}

			\def \bfS {\mathbf{S}}
	\def \calT {\mathcal{T}}		
			\def \bfU {\mathbf{U}}
	\def \calV {\mathcal{V}}		
			\def \bfW {\mathbf{W}}
			\def \bfX {\mathbf{X}}
			\def \bfY {\mathbf{Y}}
	\def \calZ {\mathcal{Z}}		\def \bfZ {\mathbf{Z}}

			\def \bfb {\mathbf{b}}

			\def \bff {\mathbf{f}}

\def \Mstar {M^{\star}}

\def \Cov {\mathrm{Cov}}

\def \conD {\overset{D}\longrightarrow}

\def \bfS {\bm{S}}
\def \bfY {\mathbf{Y}}

\def \rank {\mathrm{rank}}

\def \diag {\mathrm{diag}}

\def \init {\mathrm{init}}

\def \naive {\mathrm{naive}}
\def \full {\mathrm{full}}

\def \rest {\mathrm{rest}}

\def \proj {\mathrm{proj}}

\def \bl {\mathrm{bl}}

\newcommand{\norm}[1]{\left\Vert#1\right\Vert}

\DeclareMathOperator*{\argmin}{arg\,min}

\numberwithin{equation}{section}

\theoremstyle{definition}

\theoremstyle{plain}
\newtheorem{theorem}{Theorem}[section]

\newtheorem{assumption}{Assumption}[section]
\newtheorem*{assumption*}{Assumption}

\def \conD {\overset{D}\longrightarrow}
\def \bbE {\mathbb{E}}			\def \bfE {\mathbf{E}}

\allowdisplaybreaks

\usepackage{algorithm,algpseudocode}

\makeatletter
\newenvironment{breakablealgorithm}
{
	\begin{center}
		\refstepcounter{algorithm}
		\hrule height.8pt depth0pt \kern2pt
		\renewcommand{\caption}[2][\relax]{
			{\raggedright\textbf{\fname@algorithm~\thealgorithm} ##2\par}%
			\ifx\relax##1\relax 
			\addcontentsline{loa}{algorithm}{\protect\numberline{\thealgorithm}##2}%
			\else 
			\addcontentsline{loa}{algorithm}{\protect\numberline{\thealgorithm}##1}%
			\fi
			\kern2pt\hrule\kern2pt
		}
	}{
		\kern2pt\hrule\relax
	\end{center}
}
\makeatother

\newdateformat{mydate}{\monthname[\THEMONTH] \THEYEAR}

\begin{document}

\def\spacingset#1{\renewcommand{\baselinestretch}%
{#1}\small\normalsize} \spacingset{1}
  
\if0\blind
{ 
\title{\bf Inference for Low-rank Models without Estimating the Rank}
    \author[1]{Jungjun Choi}
  \author[2]{Hyukjun Kwon}
  \author[3]{Yuan Liao}
  \affil[1]{Department of Statistics, Columbia University}
  \affil[2]{Department of Operations Research and Financial Engineering, Princeton University}
  \affil[3]{Department of Economics, Rutgers University}
  \maketitle
} \fi

\if1\blind
{   
 \bigskip
 \bigskip
 \bigskip
\begin{center}
\spacingset{1.3}
    {\LARGE\bf Inference for Low-rank Models without Estimating the Rank}
\end{center}
   \medskip
} \fi

\bigskip
\begin{abstract}
This paper studies the inference about linear functionals of high-dimensional low-rank matrices. While most existing inference methods would require consistent estimation of the true rank, our procedure is robust to rank misspecification, making it a promising approach in applications where rank estimation can be unreliable. We estimate the low-rank spaces using pre-specified weighting matrices, known as diversified projections. A novel statistical insight is that, unlike the usual statistical wisdom that overfitting mainly introduces additional variances, the over-estimated low-rank space also gives rise to a non-negligible bias due to an implicit ridge-type regularization. We develop a new inference procedure and show that the central limit theorem holds as long as the pre-specified rank is no smaller than the true rank. In one of our applications, we study multiple testing with incomplete data in the presence of confounding factors and show that our method remains valid as long as the number of controlled confounding factors is at least as large as the true number, even when no confounding factors are present. 
\end{abstract}

\noindent%
{\it Keywords:} High-dimensional models; low-rank matrix estimation; rank misspecification; weak factors
\vfill

\newpage
\spacingset{1.9} 

\section{Introduction}\label{sec:intro}

The objective of this paper is to make inference about a  low rank matrix without consistently estimating its rank. 
We consider the following linear model:
\begin{align}
   {\bf Y=X \circ M^{\star}+E} \label{eq:model}
\end{align}
where $\bf{Y}$, $\bfX$, ${\bf\Mstar}$, and $\bfE$ are $N \times T$ matrices, with $N$ and $T$ approaching infinity, and $\circ$ represents the matrix entry-wise product. The outcome matrix $\bfY$ and the regressor matrix $\bfX$ are observed,  and $\bfE$ is the matrix of noise. The target of interest is the matrix   ${\bf\Mstar}$, whose rank (denoted by $r$)  is unknown and low compared to its dimensions.   Our model incorporates various models such as noisy matrix completion, heterogeneous treatment effects estimation, and varying coefficients models.

The inferential theory for low rank matrices has been developed in the recent   literature, as in \cite{chernozhukov2018inference, chernozhukov2023inference, chen:2019inference, xia2021statistical,yan2024inference}. Most of these methods utilize the  estimated eigenvectors for rank reductions, so we call them  principal components analysis (PCA)-based methods. A standard PCA-based  inference  procedure can be outlined as follows \citep[][]{chen:2019inference,xia2021statistical}: 
\begin{enumerate} 
    \item[Step 0:] Fix the  \textit{true} rank of ${\bf \Mstar}$ (or consistently estimate the true rank).
    \item[Step 1:] Obtain an initial estimator ${\bf \widetilde{M}^{\init}}$.
    \item[Step 2:] Subtract a  bias correction term: ${\bf \widetilde{M}}^{\naive}= {\bf\widetilde{M}}^{\init}- \calB_1.$
    \item[Step 3:] Project ${\bf\widetilde{M}}^{\naive}$ onto low-rank spaces that are consistently estimated by using PCA with the ``correct'' knowledge of the true rank.
\end{enumerate}

Here $\mathcal{B}_1$ captures a shrinkage bias one often encounters from the initial estimator. Clearly, one needs to  start by taking the true rank (or its consistent estimator), which however requires that the signal-to-noise ratio be sufficiently high. This assumption often breaks  down in finite sample applications.  In fact,  the rank estimation  is  threatened by the presence of weak factors, raising severe concerns in applications. 
For instance, in financial applications it is often the case that the first eigenvalue is much  larger than  the remaining eigenvalues, so consistent rank estimators often identify only one factor, which however violates empirical practices in   asset pricing. As another example in the forecast practice, empirical eigenvalues do not decay as fast as the theory requires, so it is often difficult to determine the   cut-off value  to separate  ``spiked eigenvalues" from the remaining ones. 

A simple   solution   is to  over-estimate the rank. To avoid the risk of under-estimating the rank, one can select a sufficiently  large  rank, and use it throughout the inference procedure. However, when the PCA-based debiasing methods are employed with the over-estimated rank, they fail to exclude the eigenvectors that are generated from and thus strongly correlated with the noise. In addition, the over-estimated eigenvectors correspond to non-spiked eigenvalues, which are  well known to be inconsistent in high dimensional settings \citep[e.g.,][]{johnstone2009consistency}.

This paper makes a novel contribution to the  low-rank inference  literature by proposing   a  procedure   robust to the rank over-estimation. In order to circumvent the aforementioned issues of the PCA-based method, we adopt a non-PCA based method, known as \textit{diversified projection} (DP) which was recently proposed by  \cite{fan2022learning}  in the pure factor model. Similar to the PCA, the DP is a dimension reduction method that projects the original high-dimensional space to a low-rank space.  But it is much  simpler  than the PCA:   it does not require calculating the eigenvalue/eigenvectors, so demands much weaker conditions on the eigen-structures of the low-rank matrix.  We employ the DP  to estimate  a ``larger'' low-rank space, which is  \textit{inconsistent}  to the true singular vector spaces.   Nevertheless, the simple structure of the DP enables us to  characterize the  over-estimated rank spaces  relatively easily.   Importantly, we do  not require correctly specifying (or consistently estimating) the true rank of the underlying matrix.

 Although our adoption of the DP approach is inspired by \cite{fan2022learning}, the primary difference lies in our objective. Their  underlying model is a pure factor model with no missing data, and the focus is exclusively on the extracted factors. In contrast, motivated by the study of treatment effects, our main goal is to make inferences about the low-rank matrix itself. When the objective shifts from focusing on factors to directly examining the low-rank matrix, a novel statistical insight emerges: a new source of bias from over-estimating the low-rank space when the rank exceeds its true value. Contrary to the conventional view that overfitting primarily introduces additional variance, we show that in the context of low-rank inference, an over-estimated low-rank space can be characterized by a \textit{Tikhonov-type function}, leading to an implicit ridge-type regularization bias.   Our statistical interpretation is that the over-estimated  low-rank space is highly correlated with the model's noise,  and such   correlation depends on  a non-stochastic second moment of the noise, giving rise to the new bias.  This issue would not arise if the objective were solely focused on the extracted factors for other types of inference.\footnote{For instance, \cite{fan2022learning} applied estimated factors to ``factor-augmented regression" problems (e.g., forecasts and post-selection inference). In such cases, additional steps often involve regressing external time series on the extracted factors, which can ``average out" the effect of over-estimating the low-rank space.}

 Our condition is   more robust to the strength of the singular values than the PCA-based approach:  we allow weak  singular values of ${\bf\Mstar}$, which just need to be  stronger than $\sqrt{T\log^2 N}$. This condition is only slightly stronger than the setting of ``weak factors" in \cite{onatski2012asymptotics} in the pure factor model, whose condition was $\sqrt{T }$. On the contrary, the usual PCA-based approaches  would require the singular values be  $\sqrt{T N^c}$ for some $c\in(1/2,1]$ \citep[see][]{bai2023approximate}.

A practical reason for allowing the true rank to be over-estimated is to account for confounding factors (CF)  in variable selection and multiple testing. Standard statistical inference can be significantly affected by strong correlations among variables due to hidden confounding factors, so needs to be adjusted. The key practical question is determining how many CF to  account for, including the special case that there are none---though this is unknown in practice, so as a precaution,   statisticians often account for some CF regardless. In one of our statistical applications, we  address this issue within the context of multiple testing. We show that our method remains valid across all bounded numbers of CF, including the special case where  none are present.
 

We recommend a few specific choices for the weighting matrices in  Section \ref{sec:choiceofW}. One of the appealing recommended choices is to use transformations of initial observations, which \text{does not} require extra data. Moreover,  we also note that   the use of  extra information is not uncommon in a wide range of literature.

  There is a large literature on estimation of high-dimensional low-rank matrices, such as  \cite{candes2009exact,keshavan2010matrix,negahban2011estimation,klopp2014noisy,cai2018rate,cape2019two,abbe2020entrywise,koltchinskii2020efficient,cai2021subspace,zhu2022high} among many others. In this literature, profound theories on optimal  rates of convergence have been developed, whereas as we commented earlier, the distributional theory has been studied in the more recent literature. \cite{barigozzi2020consistent} proposed a PCA-based method to   over-estimate the rank and achieved insightful rate results. 


We adopt the following notations. Let $\norm{\cdot}$ and $\norm{\cdot}_*$ denote the matrix operator norm and nuclear norm, respectively. Also, we use $\norm{\cdot}_{2,\infty}$ to denote the largest $l_2$ norm of all rows of a matrix. We write $\sigma_{\max}(\cdot)$ and $\sigma_{\min}(\cdot)$ to represent the largest and smallest singular values of a matrix, respectively, and $\sigma_{j}(\cdot)$ to represent the $j$th largest singular value of a matrix. For a matrix $\bfA$, define $\mathrm{span}(\bfA)$ as the linear space spanned by the columns of matrix $\bfA$. When $\bfA'\bfA$ is invertible, define $\bfP_\bfA = \bfA (\bfA'\bfA)^{-1}\bfA'.$ For a vector $\mathbf{v}$, $\diag(\mathbf{v})$ represents the diagonal matrix whose diagonal entries are $\mathbf{v}$ in order. For two sequences $a_{NT}$ and $b_{NT}$, we denote $a_{NT} \ll b_{NT}$ (or $b_{NT} \gg a_{NT}$) if $a_{NT} =o(b_{NT})$, $a_{NT} \lesssim b_{NT}$ (or $b_{NT} \gtrsim a_{NT}$) if $a_{NT} =O(b_{NT})$, and $a_{NT} \asymp b_{NT}$ if $a_{NT} \lesssim b_{NT}$ and $a_{NT} \gtrsim b_{NT}$ (almost surely if random).  Finally, due to page limits, all proofs and some simulation studies are provided in the appendix.

\section{Model and Estimation} \label{sec:modelandestimation}

In the linear model \eqref{eq:model}, we assume that the matrix $\bf\Mstar$ has a low-rank structure:
\begin{align}
    \bfY=\bfX \circ {\bf M^{\star}}+\bfE = \bfX \circ ({\bm\beta} \bfF')+ \bfE\label{eq:model2}
\end{align}
where $\bm\beta$ is an $N\times r$ matrix of  rescaled left singular vectors, and $\bfF$ is a $T\times r$  matrix of the rescaled right singular vectors; rescaled by the singular values.  Model \eqref{eq:model2} has numerous  applications including varying coefficient models, heterogeneous treatment effects \citep{athey2021matrix}, and matrix completion problems.

\subsection{Diversified projection}

The diversified projection (DP)  is a dimension reduction technique that projects a high-dimensional object onto a low-rank space. To illustrate the idea, consider the high-dimensional factor model:
\begin{align*}
    \bfY={\bf\Mstar}+\bfE= {\bm\beta} \bfF' + \bfE.
\end{align*}
To estimate $\bf\Mstar$, one approach   is to  take advantage of 
the low-rank structure, by  applying projections  as follows:
\begin{align*}
    \widehat{\bfM}=\bfP_{\widetilde{{\bm\beta}}} \bfY \bfP_{\widetilde{\bfF}}.
\end{align*}
Here, $\bfP_{\widetilde{{\bm\beta}}}$ and $\bfP_{\widetilde{\bfF}}$ are the projection matrices that estimate the true projections $\bfP_{\bm\beta}$ and $\bfP_{\bfF}$, respectively. So, $ \widehat{\bfM}$  reduces to the ``intrinsic dimension" of the parameters  by projecting the data matrix onto the low-dimensional subspaces, which are respectively spanned by $\widetilde{\bm\beta}$ and $\widetilde \bfF$. 
Usually, this is accomplished by employing PCA, where  columns of $\widetilde{\bm\beta}$ and $\widetilde \bfF$ respectively denote the    top left and right singular vectors of $\bfY$. However, one major limitation of PCA-based low-rank projection is that it requires $\rank(\widetilde{\bm\beta})$ and $\rank(\widetilde{\bfF})$  be either equal to the true rank or a consistent estimator for it, which is often a strong assumption. As we commented in the introduction, the consistent estimation of the true rank is  threatened by the strength of factors, raising a severe concern in practical applications. 

In the context of pure factor model, \cite{fan2022learning} proposed  diversified projection (DP),  as an alternative   low-rank projection.   This method begins by specifying two weighting matrices, an $N \times R$ matrix $\bfW_{\bm\beta}$ and a $T \times R$ matrix $\bfW_\bfF$, and defining:
\begin{align*}
    \widetilde{\bm\beta}= \frac{1}{T} \bfY \bfW_{\bfF} \quad \text{and} \quad \widetilde{\bfF}= \frac{1}{N}\bfY' \bfW_{\bm\beta}
\end{align*}
where the weighting matrices consist of ``diversified elements'', but not necessarily eigenvectors. These weighting matrices should satisfy: 

(i) they are uncorrelated with the noise $\bfE$.

(ii) they are correlated with the actual $\bm\beta$ and $\bfF$.

 Instead of the correct estimation of the number of factors,  it is only required that the rank of the weighting matrix, $R$, be  no smaller than the true number of factors, $r$.

Section \ref{sec:choiceofW} will give specific recommendations for choosing the weighting matrices in applications. One of the appealing recommendations  is to use transformations of initial observations, which \text{does not} require extra data. Moreover,  we also note that   the use of  extra information is not uncommon in a wide range of literature. To name a few,  \cite{fan2016projected} and   \cite{kelly2020instrumented} use firms' characteristics to study excess returns. In the matrix completion literature, the side information is widely used \citep[e.g.,][]{jain2013provable,xu2013speedup, chiang2015matrix,wang2018high}.


\subsection{Formal procedure}\label{sec:formalprocedure}

We utilize the diversified projections  in the general form of low-rank inference problems.  To begin with, we pre-determine  weighting matrices, $\bfW_{\bm\beta}$ and $\bfW_{\bfF}$, whose rank   $R$ is pre-determined, but not necessarily equal to the true rank $r$. The theory holds as long as $R\geq r$.  To account for heterogeneity, let $\widehat{\bf\Pi}=\diag(\widehat{p}_1, \ldots, \widehat{p}_N)$ where $\widehat{p}_i=T^{-1} \sum_{t=1}^T X_{it}^2$, and 
$\widehat{\bf\Psi}=N^{-1}\diag(\sum_{j=1}^N X^2_{j1}\widehat{p}_j^{-2}, \ldots, \sum_{j=1}^N X^2_{jT}\widehat{p}_j^{-2})$.

\begin{breakablealgorithm}
		\caption{\small  Formal estimation procedure}
		\label{alg:estimation}
		\begin{algorithmic}
    			\noindent \textbf{Step 1:} \textit{Initialization}. Obtain an initial   estimator $\widetilde{\bfM}^{\init}$ as to be described in Section \ref{sec:initial}, and pre-determine the weighting matrices $\bfW_{\bm\beta}$ and $\bfW_\bfF$ with rank $R$. \\
 {\textbf{Step 2:} \textit{First bias correction.}	 Let $\calB_1=\widehat{\bf\Pi}^{-1} \bfX \circ (\bfX\circ \widetilde{\bfM}^{\init}-\bfY)$. Compute \vspace{-0.7cm}$$\widetilde{\bfM}^{\naive}=\widetilde{\bfM}^{\init} - \calB_1.$$	} \\\vspace{-0.7cm}
 {\textbf{Step 3:} \textit{Diversified projections.} Let $\widetilde{\bm\beta}= T^{-1}  \widetilde{\bfM}^{\naive} \bfW_\bfF$ and $\widetilde{\bfF}= N^{-1} \widetilde{\bfM}^{\naive \prime} \bfW_{\bm\beta}.$} 
 \vspace{-0.7cm}
 $$
\widetilde{\bfM}^{\proj}=\bfP_{\widetilde{\bm\beta}} \widetilde{\bfM}^{\naive}\bfP_{\widetilde{\bfF}}.
   $$  \\
   \vspace{-0.7cm}
 \textbf{Step 4:} \textit{Second bias correction.} Compute $\widehat{\bfM}= \widetilde{\bfM}^{\proj}-\calB_2$    
   where     \vspace{-0.7cm}
   \begin{align*} 
       \calB_2=  \widetilde{\sigma}^2\left[\frac{T}{N}  \bfP_{\widetilde{\bm\beta}}\widehat{\bf\Pi}^{-1} \bfW_{\bm\beta} ( \widetilde{\bfF}'\widetilde{\bfF})^{-1}\widetilde{\bfF}' + \frac{N}{T} \widetilde{\bm\beta}(\widetilde{\bm\beta}'\widetilde{\bm\beta})^{-1} \bfW_\bfF' \widehat{\bf\Psi} \bfP_{\widetilde \bfF}\right] 
   \end{align*}  \\ 
  with  a  variance estimator $\widetilde{\sigma}^2$.\footnote{For the variance estimation, we can use $\widetilde{\bfM}^{\init}.$ Specifically, when $\bfX$ is a general regressor matrix, we define $\widetilde{\sigma}^2=(NT)^{-1}\sum_{j=1}^N \sum_{s=1}^T (Y_{js}-X_{js}\widetilde{M}^{\init}_{js})^2$.  {When $\bfX$ is binary, we may use $\widetilde{\sigma}^2=(\sum_{j=1}^N \sum_{s=1}^T X_{js})^{-1} \sum_{j=1}^N \sum_{s=1}^T X_{js}(Y_{js}-\widetilde{M}^{\init}_{js})^2$ instead.}}  
		\end{algorithmic}
	\end{breakablealgorithm}

Our Steps 1-3 are  in spirit  similar to that of the existing procedure outlined in the introduction. In particular,  the bias $\calB_1$ in Step 2 is similar to the one derived by \cite{chen:2019inference} and \cite{xia2021statistical}, who reach the following decomposition after the debias:
\begin{align*}
    \widetilde{\bfM}^{\naive} ={\bf\Mstar} +\calZ={\bf\Mstar} + \underbrace{\widehat{\bf\Pi}^{-1}\bfX   \circ \bfE}_{\text{dominant noise term}}+ \underbrace{ (\bf{1}_N\bf{1}_T'-\widehat{\bf\Pi}^{-1}\bfX \circ \bfX) \circ (\widetilde{\bfM}^{\init} - {\bf\Mstar})}_{\text{higher-order initial estimation error}}.
\end{align*}

But we also have  three key differences: First, we do not need  to specify  the true rank or its consistent estimator.  Secondly, the low-rank projections in Step 3, $\bfP_{\widetilde{\bm\beta}}$ and $\bfP_{\widetilde{\bfF}}$, are \textit{inconsistent}.  Due to the rank over-estimation, they are ``larger'' than the true low-rank projections, $\bfP_{\bm\beta}$ and $\bfP_\bfF$ (we will make this insight precise later). Finally, the inconsistency of the estimated projections gives rise to a new bias correction in Step 4, which is the main novel statistical insight of this paper. In the next subsection  we explain the source of  this new bias in details.


\subsection{A new source of bias and Tikhonov-type functions}\label{sec:tikhonov}

 After Step 2, we end up with $ \widetilde{\bfM}^{\naive}$. Write it as 
\begin{align*}
    \widetilde{\bfM}^{\naive}= {\bf\Mstar} + \calZ 
\end{align*}
where $\bf\calZ$ is the estimation error. Therefore, we have
\begin{align*} 
\bfP_{\widetilde{\bm\beta}}   \widetilde{\bfM}^{\naive} \bfP_{\widetilde{\bfF}}-\bf\Mstar = (\bfP_{\widetilde{\bm\beta}} {\bf\Mstar} \bfP_{\widetilde{\bfF}}-\bf\Mstar)+\bfP_{\widetilde{\bm\beta}} \calZ \bfP_{\widetilde{\bfF}}. 
\end{align*}

 The first term on the right hand side yields  the asymptotic normality of the estimator.
For now we will focus on the second term $\bfP_{\widetilde{\bm\beta}} \calZ \bfP_{\widetilde{\bfF}}$. Recall that  $R$ denotes the rank of $\bfP_{\widetilde{\bm\beta}}$ and $\bfP_{\widetilde{\bfF}}$. In the usual case $R=r$, this term is asymptotically negligible. But  when $R>r,$   $\mathrm{span}(\widetilde{\bm\beta})$ and $\mathrm{span}(\widetilde{\bfF})$   would also  encompass additional noise components   that are   orthogonal to $\mathrm{span}(\bm\beta)$ and $\mathrm{span}(\bfF)$, but are  strongly correlated with the noise $\calZ $. This strong correlation renders the second term asymptotically non-stochastic, introducing a bias through an implicit regularization, as we explain in detail below.


To understand the intuition,  note that the diversified projection yields an $R \times r$ rotation matrix $\bfH$ such that 
\begin{align*}
    \frac{1}{\sqrt{T}} \norm{\widetilde{\bfF}-\bfF\bfH'} = o_P(1).
\end{align*}
In the usual setting when $R=r$, the rotation matrix is well invertible (all its singular values are bounded away from zero), but this is no longer the case when $R>r$. In this case, the  following  rank-$r$ matrix becomes degenerate: 
\begin{align*}
    \bfS_\bfF\coloneqq \frac{1}{T}\bfH\bfF'\bfF\bfH', \quad R \times R.
\end{align*}
On the other hand, define  $$\bfS_{\widetilde{\bfF}}=\frac{1}{T} \widetilde{\bfF}' \widetilde{\bfF}, \quad R \times R.$$
We shall show that  when $R>r$,   $\bfS_{\widetilde{\bfF}}$ is  still asymptotically invertible, but its eigenvalues may decay very fast.  Nevertheless,  the projection matrix
$\bfP_{\widetilde \bfF}=\widetilde \bfF(\widetilde \bfF'\widetilde \bfF)^{-1}\widetilde \bfF'$ is still well defined  with probability approaching one.

The asymptotic property of the projection matrix critically depends on the following ridge-type projection function, also known as \textit{Tikhonov-regularization function}:
\begin{align*}
    f(x) \coloneqq (\bfH\frac{1}{T}\bfF' \bfF \bfH' +x \bfI_R )^{-1}.
\end{align*}
In fact,   there is a rate $x_{NT} \rightarrow 0$ such that, $\bfS_{\widetilde{\bfF}}^{-1} \approx f(x_{NT})$. The key challenge, however, is that   $f(x)$ is discontinuous at $x=0$ and $\lim_{x \rightarrow 0}f(x)$ does not exist. Therefore when $R>r$, the inverse matrix $\bfS_{\widetilde{\bfF}}^{-1}$ \textit{does not} converge in probability to  the generalized inverse $\bfS_\bfF^+$, its population counterpart.

The discontinuity challenge can be avoided in our context by considering  the following \textit{rescaled Tikhonov-regularization function}:
\begin{align*}
    \widetilde{f}(x) \coloneqq \bfH'(\bfH\frac{1}{T}\bfF' \bfF \bfH' +x \bfI_R )^{-1}\bfH.
\end{align*}
Unlike $f(x)$, the rescaled Tikhonov function is  continuous in neighborhoods of zero and has $(\frac{1}{T}\bfF'\bfF)^{-1}$ as its limit when $x \rightarrow 0$.   Fortunately, in  the low-rank inference problem, it is sufficient to study the behavior of $\widetilde{f}(x) $ instead of $f(x)$ because the projection matrix $\bfP_{\widetilde{\bfF}}$ asymptotically depends on $\bfS_{\widetilde{\bfF}}$ through $\bfH'\bfS_{\widetilde{\bfF}}^{-1}\bfH$. Let $\bfS_\bfF^{+}$ denote the generalized inverse of $\bfS_\bfF$. We shall show that while $\|\bfS^{-1}_{\widetilde{\bfF}}-\bfS^+_{\bfF}\| \neq o_P(1),$ when rescaled  by $\bfH$, we have,  
    \begin{align*}
        \left\|\bfH'\bfS^{-1}_{\widetilde{\bfF}} \bfH- \left(\frac{1}{T}\bfF'\bfF\right)^{-1}\right\|
        \approx \| \widetilde f(x_{NT}) -\widetilde f(0)\| 
        =O_P( x^2_{NT})
    \end{align*}
    for some sequence $x_{NT}\rightarrow 0$. 

    Above all, when $R>r$, the correlation between   the estimated low-rank projection matrices and $ \calZ $ can be 
    characterized by a Tikhonov-type regularization  function,  akin to ridge regression, which acts as an implicit regularization. 
   Therefore, a key new statistical insight of this paper is that, unlike the conventional understanding where overfitting primarily results in additional variance, in the context of low-rank inference,  over-estimating the rank introduces  a novel source of  asymptotic bias through  $\bfP_{\widetilde{\bm\beta}} \calZ \bfP_{\widetilde{\bfF}}$. We identify and address this bias and reach the final estimator:
    \begin{align*}
    \widehat{\bfM}:=\widetilde{\bfM}^{\proj} - \calB_2 = \bfP_{\widetilde{\bm\beta}} {\bf\Mstar} \bfP_{\widetilde{\bfF}} + \underbrace{\bfP_{\widetilde{\bm\beta}} \calZ \bfP_{\widetilde{\bfF}}- \calB_2}_{\text{entry-wise negligible}}. 
\end{align*}
This is the motivation for introducing $\calB_2$ in Step 4.




\section{Asymptotic Results}\label{sec:asympresults}

The objective is to establish the asymptotic normality of our estimator for the group average, given by $|\calG|^{-1} \sum_{(i,t) \in \calG} \widehat{M}_{it}$, where $\calG \subset \{1, \ldots, N\} \times \{1, \ldots, T\}$ represents the group we are interested in. 

The following assumption formalizes the data generating process (DGP) of the noise $\bfE$.

\begin{assumption}[DGP for $\bfE$]\label{asp:dgpnoise}
 \begin{enumerate}
     \item[(i)] Conditioning on $({\bm\beta}, \bfF, \bfW_{\bm\beta}, \bfW_\bfF)$, $E_{it}$ is an i.i.d. (across $i$ and $t$) sub-Gaussian random variable with zero mean, a finite variance, $\sigma^2$, and sub-Gaussian norm at most $C \sigma$ for some $C>0.$ Also, $\bfE$ is independent of $\bfX$.
     \item [(ii)] When $\bfX$ is a binary matrix taking values in $\{0,1\}$, then  condition (i) may be replaced by:  There is a noise matrix $\bfE^{\star}$ that satisfies (i), and $\bfE$ can be written as $\bfE=\bfX\circ \bfE^{\star}$.
 \end{enumerate}
\end{assumption}
We note that Assumption \ref{asp:dgpnoise} (ii) accommodates the noisy matrix completion problem where $X_{it}$ indicates if the entry $(i,t)$ is observed.

The next assumption specifies the DGP of the regressor matrix $\bfX$. We allow heterogeneity in $\bfX$ across units. In the matrix completion application, it can accommodate the heterogeneous missing probabilities across $i$. Also, we allow cross-sectional weak dependence in $\bfX$ through the cluster structure, where the size of the largest cluster is allowed to grow. Let $\calC_1,\ldots, \calC_{\rho} \subset \{1, \ldots, N\}$ be non-empty and disjoint clusters such that $\cup_{g=1}^\rho \calC_g=\{1, \ldots, N\}$. We denote $\vartheta \coloneqq \max_{g=1, \ldots, \rho}|\calC_g|$.

\begin{assumption}[DGP for $\bfX$]\label{asp:dgpX}
\begin{enumerate}
    \item[(i)] Conditioning on $({\bm\beta}, \bfF, \bfW_{\bm\beta}, \bfW_\bfF)$, $X_{it}$ is i.i.d. across $t$ for each $i.$  In addition, $X_{it}$ are   independent across clusters, and they are allowed to be dependent within clusters. Overall, 

    $\max_{t \leq T} \max_{j \leq N}\sum_{i=1}^N \left| \Cov(X^2_{it}, X^2_{jt}|{\bm\beta}, \bfF, \bfW_{\bm\beta}, \bfW_\bfF) \right| \leq  C$ for some $C>0.$
    \item[(ii)] Let $p_i \coloneqq \bbE[X_{it}^2|{\bm\beta}, \bfF, \bfW_{\bm\beta}, \bfW_\bfF]$ and $p_{\min}\coloneqq \min_{i \leq N}p_i$. We assume that $p_{\min}$ is bounded away from zero.
    \item[(iii)]   $\max_{i\leq N}\max_{t \leq T} X^2_{it} < C$ for some $C>0$ almost surely.
\end{enumerate}
\end{assumption}

Next, the following assumption specifies the class of high-dimensional matrices we are interested in. We require $\bf\Mstar$ to be of low-rank and incoherent.

\begin{assumption}[Structure of $\bf\Mstar$]\label{asp:incoherence}
\begin{enumerate}
    \item[(i)] Low-rank factor structure: We assume $\bf\Mstar={\bm\beta} \bfF'$ with an $N \times r$ loading matrix $\bm\beta$ and a $T \times r$ factor matrix $\bfF$. We assume that $r$ is bounded.
    \item[(ii)] Incoherence: The matrix $\bf\Mstar$ is incoherent in that:
    \begin{align*}
    \norm{\bm\beta}_{2,\infty} \lesssim \frac{\sigma_{\max}({\bm\beta})}{\sqrt{N}} \quad \text{and} \quad \norm{\bfF}_{2,\infty} \lesssim \frac{\sigma_{\max}({\bfF})}{\sqrt{T}}.
\end{align*}
     \item[(iii)] $T/N \rightarrow C$ for some $C \in (0,\infty)$.
\end{enumerate}
\end{assumption}

The following assumption specifies conditions on the diversified weights and the strength of factors. Regarding the factor strength, without loss of generality, we fix $\sigma_{\min}(\bfF) \asymp \sigma_{\max}(\bfF) \asymp \sqrt{T}$ and accommodate the weak factor by allowing the aggregated factor loadings to be ``small."

\begin{assumption}[Diversified weights and weak factors]\label{asp:diversifiedweightsandweakfactor} 
We define $\bfH_{\bm\beta}\coloneqq N^{-1} \bfW_{\bm\beta}'{\bm\beta}$ and $\bfH_\bfF\coloneqq T^{-1}\bfW_\bfF' \bfF$.
\begin{enumerate}
    \item[(i)] $\bfW_{\bm\beta}$ is an $N \times R$ matrix and $\bfW_\bfF$ is a $T \times R$ matrix, where $R$ is bounded and $R \geq r$. $\bfW_{\bm\beta}$ and $\bfW_\bfF$ are independent of $\bfE.$\footnote{When Assumption \ref{asp:dgpnoise} (ii) holds with $\bfE=\bfX \circ \bfE^{\star}$, we assume that $\bfW_{\bm\beta}$ and $\bfW_\bfF$ are independent of $\bfE^{\star}.$} Also, almost surely, $\max\{\norm{\bfW_{\bm\beta}}_{2, \infty}, \norm{\bfW_\bfF}_{2, \infty}\}<C$, $\min\{\sigma_{R}(N^{-1} \bfW_{\bm\beta}'\bfW_{\bm\beta}), \sigma_{R}(T^{-1} \bfW_\bfF'\bfW_\bfF) \}>c$ for some $c, C>0$, and, the ranks of $\bfH_{\bm\beta}$ and $\bfH_{\bfF}$ are $r$.
    \item[(ii)] $\log N \ll \sigma_{\min}({\bm\beta}) \asymp \sigma_{\max}({\bm\beta}) \lesssim \sqrt{N}$ and $\sigma_{\min}(\bfH_{\bm\beta}) \asymp \sigma_{\max}(\bfH_{\bm\beta}) \asymp \sigma_{\max}({\bm\beta})/ \sqrt{N} $.
    \item[(iv)]$\sigma_{\min}(\bfH_\bfF) \asymp \sigma_{\max}(\bfH_\bfF) \asymp C$ for some constant $C>0$.
\end{enumerate}
\end{assumption}

Our condition on the factor strength is relatively weak. For example, suppose that there exist a bounded sequence $a_N=O(1)$, an $N \times r$ matrix ${\bm\beta}_0$, and diversified weights $\bfW_{\bm\beta}$ such that:
\begin{align*}
    {\bm\beta} = a_N   {\bm\beta}_0 \quad \text{and} \quad \sigma_{\min}\left(\frac{1}{N}\bfW_{\bm\beta}' {\bm\beta}_0 \right) \asymp \sigma_{\max}\left(\frac{1}{N}\bfW_{\bm\beta}' {\bm\beta}_0 \right)\asymp C
\end{align*}
for some constant $C>0$. Here, we assume $\sigma_{\min}(N^{-1}{\bm\beta}_0'{\bm\beta}_0) \asymp \sigma_{\max}(N^{-1}{\bm\beta}_0'{\bm\beta}_0)\asymp C$  for some constant $C>0$, so that ${\bm\beta}_0$ can be regarded as the ``standardized direction'' of ${\bm\beta}$. Therefore, the strength of ${\bm\beta}$ is governed by the sequence $a_N$. Then, Assumption \ref{asp:diversifiedweightsandweakfactor} (ii) and (iii)  can be simplified to:
 $  \log N \ll \sqrt{N} a_N. $
This implies   the condition on the  factor strength as:
$$\sigma_{\min}\left(  {\bm\beta}'{\bm\beta}\right) \asymp Na_N^2 \gg \log^2 N ,$$
which is   only slightly stronger than the definition of ``weak factors'' in \cite{onatski2012asymptotics} in the pure factor model. In contrast, the usual requirement for inference in the PCA-based approaches \citep[see][]{bai2023approximate} corresponds to $\sigma_{\min}\left( {\bm\beta}'{\bm\beta}\right)\gtrsim N^{c}$ for some $c \in (1/2,1].$

We now introduce notations for groups and assumptions for them. This paper explores three distinct types of group averages of $\bf\Mstar$ for inference: i) block averages, ii) serial averages, and iii) cross-sectional averages. For the block, denoted as $\calG_{\bl}$, we define $\calG_{\bl} = \calI \times \calT$ where $\calI \subset \{1, \ldots, N\}$ and $\calT \subset \{1, \ldots, T\}$. Similarly, for cross-sectional groups, denoted as $\calG_{\mathrm{cs}}$, let $\calG_{\mathrm{cs}} = \{1, \ldots, N\} \times \calT$ where $\calT \subset \{1, \ldots, T\}$, and for serial groups, denoted as $\calG_{\mathrm{serial}}$, let $\calG_{\mathrm{serial}} = \calI \times \{1, \ldots, T\}$ where $\calI \subset \{1, \ldots, N\}$. For brevity, we present our main results for the block average case only. The results for other cases, which are similar, are provided in the appendix.

\begin{assumption}[Block shape]\label{asp:blocksize}
We assume i) $\max\{|\calI|^2 \vartheta^{3} \log^{3}N,|\calI| \vartheta^{6} \log^{6}N \}  =o(N)$, ii) $\max\{|\calT|^2 \vartheta^{3} \log^{3}T,|\calT| \vartheta^{6} \log^{6}T \}  = o(N)$, and iii) $\sqrt{\max\{|\calI|, |\calT|\}}\log N  \ll \sigma_{\min} ({\bm\beta})$.
\end{assumption}

Finally, it is critical for the initial estimator $\widetilde{\bfM}^{\init}$ to have desirable properties, such as reasonable convergence rates and weak correlation with the noise. We provide the construction of such $\widetilde{\bfM}^{\init}$, that is based on the nuclear norm penalized estimation, in Section \ref{sec:initial}.

Define the covariance estimator:
\begin{align*}
    \widehat{\calV}_{\calG_{\bl}}
    \coloneqq \frac{\widetilde{\sigma}^2}{|\calT|^2 N^2}\sum_{t \in \calT} \sum_{j=1}^N  (\widehat{\bfM}_{\calI, \cdot} \widetilde{\bfF}(\widetilde{\bfF}'\widetilde{\bfF})^{-1} \bfW_{{\bm\beta},j } \widehat{p}_{j}^{-1}X_{jt})^2 + \frac{\widetilde{\sigma}^2}{|\calI|^2 T^2}\sum_{i \in \calI} \sum_{s=1}^T (\widehat{\bfM}_{\cdot, \calT}'\widetilde{\bm\beta}(\widetilde{\bm\beta}'\widetilde{\bm\beta})^{-1}\bfW_{\bfF,s } \widehat{p}_i^{-1} X_{is} )^2
\end{align*}    
where $\widehat{\bfM}_{\calI, \cdot}\coloneqq |\calI|^{-1} \sum_{i \in \calI}\widehat{\bfM}_{i, \cdot}$, and $\widehat{\bfM}_{\cdot, \calT}\coloneqq|\calT|^{-1} \sum_{t \in \calT}\widehat{\bfM}_{\cdot,t}$.\footnote{For a matrix $\bfA$, let $\bfA_{j,\cdot}$ and $\bfA_{\cdot, j}$ denote $j$th row and $j$th column of $\bfA$, respectively.}

\begin{theorem}[Feasible CLT for block average]\label{thm:feasibleclt}
Suppose $R \geq r$ and Assumption \ref{asp:dgpnoise}-\ref{asp:blocksize} hold. In addition, the initial estimator  $\widetilde{\bfM}^{\init}$ is as constructed in Section \ref{sec:initial}. Assume that $\| |\calI|^{-1}\sum_{i \in \calI} {\bm\beta}_{i }\|> C   \sigma_{\min}({\bm\beta})/\sqrt{N}$ and $\||\calT|^{-1} \sum_{t \in \calT} \bfF_{t }\|>C$ for some $C>0.$ Then, we have
   \vspace{-0.7cm}
\begin{align*}
\widehat{\calV}_{\calG_{\bl}}^{-\frac{1}{2}} \frac{1}{|\calG_{\bl}|}\sum_{(i,t) \in \calG_{\bl}}(\widehat{M}_{it}-\Mstar_{it}) \conD \calN (0,1).
\end{align*}

\end{theorem}

The asymptotic variance depends on the choice of the DP weighting matrices. In the special case that $R$ is a consistent estimator of the true rank, these weights can be chosen optimally as the usual eigenvectors. Then the asymptotic variance attains the efficiency bound as achieved by   \cite{chernozhukov2023inference,choi2023inference}.  But more general weighting matrices will lead to   efficiency loss, which is the cost of being flexible of choosing these weighting matrices to be robust to over-estimating the rank.


\section{Construction of the Initial Estimator}\label{sec:initial}

We now formally characterize the initial estimator $\widetilde{\bfM}^{\init}$  in Step 1. 
The central limit theorem arises from the entries of $\bfP_{\widetilde{\bm\beta}} {\bf\Mstar} \bfP_{\widetilde{\bfF}}-{\bf\Mstar}$, which can be shown as weighted row and column sums of the noise matrix $\bfE$.  Not so surprisingly, however, the  weights  in these sums are correlated with $(\widetilde{\bm\beta}'\widetilde{\bm\beta})^{-1}$ and $(\widetilde{\bfF}'\widetilde{\bfF})^{-1}$ which, by construction, depend on  the initial estimator. To establish the CLT, a   technical challenge  arises from the correlation between $\widetilde{\bfM}^{\init}$ and the rows and the columns of $\bfE$.  Hence the initial estimator should be constructed in a way such that the correlation can be well controlled. 

We adopt a standard approach  that artificially creates independence that is inspired by the  sample splitting idea.   Specifically, $\widetilde{\bfM}^{\init}$ is constructed through two steps, the first step with the full sample and the second step with restricted samples.   

\noindent\textbf{Full sample.} We define  
\begin{align}
  \widetilde{\bfM}^{\full} \coloneqq  \argmin_{\bfM \in \mathbb{R}^{N \times T}} \frac{1}{2} \sum_{j=1}^N \sum_{s=1}^T \widehat{p}_j^{-1}(Y_{js}-X_{js}M_{js})^2 + \lambda \norm{\bfM}_* \label{eq:full}
\end{align}
where $\lambda>0$ is a tuning parameter.  The objective function incorporates inverse weights, $\widehat{p}^{-1}_j=(T^{-1} \sum_{t=1}^T X_{jt}^2)^{-1}$ for $j=1, \ldots, N$, to accommodate the heterogeneity in $\bfX$. Similar weighting techniques have been employed in previous works  such as \cite{ma2019missing} and \cite{choi2023inference}.

\noindent\textbf{Restricted sample.}  
 Next, to remove the  effect of correlations between  $\widetilde{\bfM}^{\full}$  and entries in $\bfE$,  we then  replace the majority  of entries in $\widetilde{\bfM}^{\full}$ with the estimates that are independent of the noises appearing in the weighted sum.  This section will focus on the block group $\calG_{\bl}$, where we are interested in making inference for the group average over $i\in \mathcal I$ and $t\in\mathcal T$:
$$
\frac{1}{|\calT||\calI|}\sum_{i\in\calI}\sum_{t\in\calT} \Mstar_{it},
$$
under Assumption \ref{asp:blocksize}.\footnote{The construction of $\widetilde{\bfM}^{\init}$ slightly differs for other group types. Refer to the appendix for other types.}    Compute  the nuclear norm penalized estimation using the sample  \textit{outside of the group}: $i\notin \mathcal I$ and $t\notin\mathcal T$, 
\begin{align*}
\widetilde{\bfM}^{\rest}:=  \argmin_{\bfM  } \frac{1}{2} \sum_{i\notin\calI} \sum_{t\notin\calT} \widehat{p}_i^{-1}(Y_{it}-X_{it}M_{it})^2 + \lambda^{\rest} \norm{\bfM}_*. 
\end{align*}

\noindent\textbf{Merging two estimates.} Then, we define the   initial estimator $\widetilde{\bfM}^{\init}$ as
\begin{align*}
\widetilde{M}^{\init}_{it}=
\begin{cases}
\widetilde{M}^{\rest}_{it} &  \text{if $i\notin \calI$ and $t\notin\calT$,}  \cr
 \widetilde{M}^{\full}_{it} & \text{otherwise.} 
\end{cases}
\end{align*}
 In the block group case, the weighted sum of noises, that leads to asymptotic normality, will consist of  rows  $i\in\calI$ and columns $t\in\calT$ of $\bfE$. By construction, the majority of entries   ($i\notin \mathcal I$ and $t\notin\mathcal T$)  in $\widetilde{\bfM}^{\init}$ are independent of these noises, as intended.
 


 
\section{Statistical Applications}\label{sec:application}

We present two specific statistical applications:  treatment effect estimation and multiple testing. In both applications the true rank is typically unknown and it is critical to develop inferential methods that are robust to over-specifying the rank.

\subsection{Heterogeneous Treatment Effects Estimation}\label{sec:treatmenteffect}

This section elaborates on the application of our theory to heterogeneous treatment effect estimation and presents formal asymptotic results. Following the causal inference literature, e.g., \cite{rubin:1974, imbens:2015}, we assume that, for each $(i,t),$ there exist two \textit{potential} outcomes, $Z^{(0)}_{it}$ and $Z^{(1)}_{it}$, where $Z^{(0)}_{it}$ represents the outcome that would be observed if $(i,t)$ is controlled and $Z^{(1)}_{it}$ represents the outcome that would be observed when $(i,t)$ is treated. We observe only one of the potential outcomes for each $(i,t)$, which basically defines two incomplete matrices.

We assume that the two potential outcomes have the following structure:
\begin{align*}
   Z^{(\iota)}_{it}= M^{(\iota)}_{it}+ E^{\star}_{it} = {\bm\beta}_{i}' \bfF_{t}^{(\iota)} + E^{\star}_{it},
\end{align*}
for each $\iota \in \{0,1\}$. By defining $D_{it}=\mathbf{1}\{\text{$(i,t)$ is treated}\}$, $\bfX^{(0)}=[1-D_{it}]_{i \leq N, t \leq T}$, and $\bfX^{(1)}=[D_{it}]_{i \leq N, t \leq T}$, we can represent the two sets of observed data in the following way:
 \begin{align}
      \bfY^{(\iota)}= \bfX^{(\iota)} \circ \bfZ^{ (\iota)} = \bfX^{(\iota)} \circ \bfM^{(\iota)} +\bfX^{(\iota)} \circ \bfE^{\star} = \bfX^{(\iota)} \circ ({\bm\beta} \bfF^{(\iota) \prime}) + \underbrace{\bfX^{(\iota)} \circ \bfE^{\star}}_{\coloneqq \bfE^{(\iota)}}, \label{eq:treatment}
 \end{align}
 for each $\iota \in \{0,1\}$. Therefore, by applying the matrix completion method to each of $\bfY^{(0)}$ and $\bfY^{(1)}$, we will obtain $\widehat{\bfM}^{(0)}$ and $\widehat{\bfM}^{(1)}$ that estimate $\bfM^{(0)}$ and $\bfM^{(1)}$ respectively.

Our goal is to perform inference about group average treatment effects for group $\calG$. We denote the treatment effect for each $(i,t)$ as $\Gamma_{it}= M^{(1)}_{it}-M^{(0)}_{it}$. Then, the group average treatment effect is defined as
\begin{align*}
    \frac{1}{|\calG|} \sum_{(i,t) \in \calG} \Gamma_{it} = \frac{1}{|\calG|} \sum_{(i,t) \in \calG} (M^{(1)}_{it}-M^{(0)}_{it}). 
\end{align*}
Then, a natural choice for the estimator for the group treatment effect would be
\begin{align*}
    \frac{1}{|\calG|} \sum_{(i,t) \in \calG} \widehat{\Gamma}_{it} = \frac{1}{|\calG|} \sum_{(i,t) \in \calG} (\widehat{M}_{it}^{(1)}-\widehat{M}_{it}^{(0)}). 
\end{align*}

Assuming that the assumptions in Section \ref{sec:asympresults} hold for each superscript $(0)$ and $(1)$, we can establish the asymptotic normality for $|\calG|^{-1} \sum_{(i,t) \in \calG} \widehat{\Gamma}_{it}- |\calG|^{-1} \sum_{(i,t) \in \calG} \Gamma_{it}$.   As in Section \ref{sec:asympresults}, we provide the result for the block average (the heterogeneous treatment effect). The results for other group types are provided in the appendix. 


We define, for each $\iota=0,1,$
\begin{align*}
    \widehat{\calV}_{\calG_{\bl}}^{(\iota)}
    &\coloneqq \frac{\widetilde{\sigma}^2}{|\calT|^2 N^2}\sum_{t \in \calT} \sum_{j=1}^N  (\widehat{\bfM}^{(\iota)}_{\calI, \cdot} \widetilde{\bfF}^{(\iota)}(\widetilde{\bfF}^{(\iota) \prime}\widetilde{\bfF}^{(\iota)})^{-1} \bfW_{{\bm\beta},j } (\widehat{p}^{(\iota)}_{j})^{-1}X^{(\iota)}_{jt})^2 \\
    &\quad + \frac{\widetilde{\sigma}^2}{|\calI|^2 T^2}\sum_{i \in \calI} \sum_{s=1}^T (\widehat{\bfM}^{(\iota) \prime}_{\cdot, \calT}\widetilde{\bm\beta}^{(\iota) }(\widetilde{\bm\beta}^{(\iota) \prime}\widetilde{\bm\beta}^{(\iota)})^{-1}\bfW_{\bfF,s }^{(\iota)} (\widehat{p}^{(\iota)}_i)^{-1} X^{(\iota)}_{is} )^2
\end{align*}    
with $\widehat{\bfM}_{\calI, \cdot}^{(\iota)}\coloneqq |\calI|^{-1} \sum_{i \in \calI}\widehat{\bfM}_{i, \cdot}^{(\iota)}$, and $\widehat{\bfM}^{(\iota)}_{\cdot, \calT}\coloneqq|\calT|^{-1} \sum_{t \in \calT}\widehat{\bfM}^{(\iota)}_{\cdot,t}$.

\begin{theorem}[Feasible CLT for heterogeneous treatment effect]\label{thm:feasibleclt-treat}
 Suppose $R \geq r$ and the assumption in Theorem \ref{thm:feasibleclt} hold for both $(0)$ and $(1).$ Also, the initial estimators $\widetilde{\bfM}^{\init, (\iota)}$, $\iota=0,1,$ are as constructed in Section \ref{sec:initial}. Then,
\begin{align*}
(\widehat{\calV}_{\calG_{\bl}}^{(0)}+\widehat{\calV}^{(1)}_{\calG_{\bl}})^{-\frac{1}{2}} \frac{1}{|\calG_{\bl}|}\sum_{(i,t) \in \calG_{\bl}}(\widehat{\Gamma}_{it}-\Gamma_{it}) \conD \calN (0,1). 
\end{align*}
\end{theorem}

\subsection{ Multiple testing with incomplete data}

In large-scale multiple testing, it is crucial to address the well-known confounding factors and the resulting strong correlations \citep[e.g.][]{leek2008general, friguet2009factor, wang2017confounder,fan2019farmtest}. We consider the model, for $t=1,...,T$:
\begin{align*}
    \bfZ_t=  {\bm\mu} + \bfU_t, \quad \text{where} \quad  \bfU_t={\bm\beta \bfF_t} + \bfE^{\star}_t.
\end{align*}
Here ${\bm\mu}= (\mu_1, \ldots, \mu_N)'$ is the mean vector and $\mathbb E \bfU_t=0$. The noise $  \bfU_t$ consists of the independent part $\bfE^{\star}_t$ and the confounding factor part ${\bm\beta \bfF_t} $.

We consider a practical situation where data is not fully observable. Let $\bfX_t$ be an $N$-dimensional vector, whose element $X_{it}=\mathbf{1}\{Z_{it}  \text{ is observed}\}$. Then, the observed data is $\bfY_t:=\bfX_t \circ \bfZ_t$, satisfying: 
$$
    \bfY_t= \bfX_t \circ ({\bm\mu} + \bfU_t).
 $$
 The objective is to test $N$ hypotheses:
\begin{align*}
    H_0^i: \mu_i = 0, \quad i=1, \ldots, N.
\end{align*}
This model differs from the usual multiple testing model in two ways: (i)  the noise  in $\bfU_t$ are strongly dependent due to the presence of confounding factors $\bm\beta \bfF_t$; and (ii) the ``data" is observed subjected to missing values, indicated by the binary vector $\bfX_t$.  

While the importance of addressing confounding correlations in multiple testing has been widely recognized \citep[e.g.,][]{wang2017confounder}, a critical yet unresolved question remains: How much confounding correlation should be accounted for? This question raises two concerns, both of which are highly relevant to practical applications.

First, researchers have relied on consistent estimation of the rank in $\bfU_t$, though ensuring its accuracy has always been a concern.  Secondly, methods that explicitly allow $\bm\beta \bfF_t$ inherently assume the presence of at least one confounding factor. But what happens if we account for confounding factors when, in reality, there are none? This corresponds to a very special case of over-estimating the rank,  where $r=0$ but $R>0$.
The fact that whether  $r$ equals zero is unknown in practice,  As a precaution, statisticians often account for  $R>0$ ``factors" regardless. In this subsection, we prove that the results are uniformly valid for all cases where $R \geq r$, even when $r=0$. This finding is empirically significant: one should always account for potential confounding correlations, as doing so does no harm even when none exist, at least asymptotically. \footnote{An alternative practice is to pretest whether $\bm\beta \bfF_t$ exists. But the power of such tests are inherently affected by the strength of the factors, which may not be detectable if factors are weak in finite sample. } 
 
 We define $\bar{Y}_i=|\calT_i|^{-1} \sum_{t \in \calT_i} Y_{it}$ where $\calT_i$ is the set of observed indices in the $i$th row. We also define the demeaned data $Y^{\mathrm{d}}_{it}= Y_{it}- \bar{Y}_i$ if $X_{it}=1$, and $Y^{\mathrm{d}}_{it}=0$ otherwise. Let $(\bfY^{\mathrm{d}},\bfX, \bfE^{\star}) $ be  the matrices of $(Y_{it}^{\mathrm{d}}, X_{it}, E_{it}^{\star})$. Then 
 $
\bfY^{\mathrm{d}}\approx \bfX\circ\bfM^{\star} +\bfE
 $, 
 where $\bfM^{\star} $  denotes the matrix of $\bm\beta \bfF_t$ and $\bfE \coloneqq \bfX\circ \bfE^{\star}$. We implement Algorithm \ref{alg:estimation} on $\bfY^{\mathrm{d}}$ to obtain $\widehat{\bfM}.$\footnote{The  initialization in Section \ref{sec:initial}, though necessary for the CLT for $\bfM^{\star}$, is not required for  testing  $\bm\mu$. Instead, we simply implement the full-sample estimation \eqref{eq:full} on $\bfY^{\mathrm{d}}$ just once and use it as $\widetilde{\bfM}^{\init}$.}  Our proposed estimator for ${\bm\mu}$ is 
\begin{align*}
    \widehat{\mu}_i&= \bar{Y}_i-\frac{1}{|\calT_i|} \sum_{t \in \calT_i} \widehat{M}_{it} \quad \text{for $i =1, \ldots, N$}.
\end{align*}


 
 \begin{theorem}\label{thm:multipletesting}
    Suppose Assumption \ref{asp:dgpnoise} (ii), \ref{asp:dgpX} hold. In addition, suppose that $\min_i |\calT_i| > \epsilon T$ for some $\epsilon>0$ almost surely. Assume the following:
    \begin{enumerate}
        \item[(i)] When $r>0$, in addition to Assumption \ref{asp:incoherence}, \ref{asp:diversifiedweightsandweakfactor}, we assume $\vartheta^6 \log^7 N \ll N$, and $\log^{\frac{3}{2}}N \ll \sigma_{\min}({\bm\beta})$ hold. Also, $\mathbb E\bfF_t=\bf{0} $ and $\{\bfF_t\}_{t \leq T}$ is i.i.d.
        \item[(ii)] When $r=0$, $\bfW_{\bm\beta}$ and $\bfW_\bfF$ are independent of $\bfE^{\star},$ and $\max\{\norm{\bfW_{\bm\beta}}_{2, \infty}, \norm{\bfW_\bfF}_{2, \infty}\}<C$, $\min\{\sigma_{R}(N^{-1} \bfW_{\bm\beta}'\bfW_{\bm\beta}), \sigma_{R}(T^{-1} \bfW_\bfF'\bfW_\bfF) \}>c$ for some $c, C>0$ almost surely. Also, $T/N \rightarrow C$ for some $C \in (0, \infty)$.
    \end{enumerate}
    Then, uniformly for $i=1, \ldots, N$ and all bounded $R\geq r \geq 0$, we have 
    \begin{align}
    \widehat{\mu}_i-\mu_i=\frac{1}{|\calT_i|} \sum_{t \in \calT_i} E_{it}+o_P\left(\sqrt{\frac{1}{|\calT_i|\log N}}\right). \label{eq:FDRresult}
\end{align}
\end{theorem}

The above expansion \eqref{eq:FDRresult} shows that the leading terms in the expansion of  $\widehat{\mu}_i-\mu_i$ are cross-sectionally weakly correlated, e.g., the confounding correlations have been successfully removed, regardless of whether the confounding factors are present. From here, one can apply the standard multiple testing approach, such as \cite{benjamini1995controlling} (B-H procedure). The standard B-H procedure implements the test as follows:  let $p_{(1)}\leq...\leq p_{(N)}$ denote the sorted p-values for each test. Then $H_0^i$ is rejected if $p_i\leq p_{(k)}$ where $k=\max\{i\leq N: p_{(i)}\leq \tau i/N\}$.  Building on (\ref{eq:FDRresult}), \cite{liu2014phase} showed that the false discovery rate of the B-H procedure can be controlled below $\tau$ asymptotically.


\section{Choices of Diversified Weights}\label{sec:choiceofW}

In this section, several choices of the diversified weighting matrices are proposed.  We emphasize that the requirement for the constructed weights is quite mild: they do not need to consistently estimate the true parameters. Instead, it suffices that they are \textit{informative} with respect to the underlying matrix parameter.

\subsection{Observed characteristics} \label{subsec:obschr}
Suppose $\beta_{ik}= g_k(\bfb_i,\eta_{ik})$ where $g_k(\cdot)$ are unknown functions,  $\bfb_i$ is a vector of observable characteristics, and $\eta_{ik}$ is noise. We define $\bfW_{\bm\beta}$ as the transformations of $\bfb_i$, i.e,. $W_{{\bm\beta}, ik}=\phi_k(\bfb_i)$  with   transformation functions $\phi_k(\cdot)$. Similarly, we assume that there  are  observable characteristics $\bff_t$   for factors. Namely, $F_{tk}=h_k(\bff_t, \xi_{tk})$,   where $\xi_{tk}$ is noise.  We then define $W_{\bfF,tk}=\varphi_k(\bff_t)$  with a set of transformation functions $\varphi_k(\cdot).$ 

It is not uncommon to observe individual-specific characteristics, which bring additional information regarding the singular vectors. For example, the ``Netflix Challenge'' called for a matrix completion problem \citep{bennett2007netflix}, where $\bfb_i$ denotes customers' demographic characteristics such as age and sex, which might be related to their preferences. In addition, films are classified according to their genres: action, romance, sci-fi, and drama, which are denoted by $\bff_t$. For example, \cite{harper2015movielens} provide user ratings of movies along with characteristics of users and movies.\footnote{Additionally, Yale University's library has documented over 40 film genres, styles, categories, and series in its Film Studies Research Guide. For further reference, see \url{https://guides.library.yale.edu/c.php?g=295800&p=1975072}.}
In financial applications such as asset pricing, $\bfb_i$ may consist of firm characteristics as numerous studies \citep[e.g.,][]{connor2012efficient, fan2016projected}. Also, observed macroeconomic factors or Fama-French factors \citep{fama1993common} can be used as $\bff_t$.  

\subsection{Exploiting extra samples}\label{subsec:extramsample}
Suppose we have access to additional data 
 $(Y_{it},X_{it})$ for  $i\in \mathcal I_1$ and $t\in\mathcal T_1$, where $\mathcal I_1$ and $\mathcal T_1$ are the index sets for extra data.
 
 Consider the subsample on $t\in\mathcal T_1$. We can write the model as 
 $
 \bfY_{t} = \bfX_{t} \circ ({\bm\beta} \bfF_t) + \text{noise},
 $  for  $t\in\mathcal T_1$, 
 where $\bfY_t$ and $\bfX_t$ are the vectors of the same subjects but observed at the  extra time $t\in \mathcal T_1$. This provides extra information about the factor loading ${\bm\beta}$, and the size of $\mathcal T_1$   is sufficient as long as it is larger than $R$.  The diversified weight $\bfW_{\bm\beta}$ can be constructed based on:  
 $$
 \bar{b}_i= \sum_{t\in\mathcal T_1} X_{it}Y_{it}\slash  \sum_{t\in\mathcal T_1} X_{it}^2.
 $$ 
  We then construct $W_{{\bm\beta},ik}= \phi_{k}(\bar{b}_i)$ for $i \leq N$ and $k \leq R$ using the nonlinear transformations of $\bar{b}_i$ to span $R$-dimensional space. Similarly, we can construct $\bfW_\bfF$ using the  additional sample   $i\in\mathcal I_1$.

\subsection{Initial transformation}\label{sec:initialtransformation}
If neither   characteristics nor the extra sample are available, we can still implement DP via transformations of the initial observations. Appealingly, it does not require extra data.

We suppose there are ``initial observations" to satisfy the following conditions:
\begin{enumerate}
    \item[(i)] For $t_0=1$, $\{E_{i,t_0}: \forall i  \}$    is independent of $\{E_{i,t}: \forall i\}$ for all  $t\geq 2$; and 
    \item[(ii)] There is a known individual $i$, say $i=1$, so that the noise $\{E_{1,t}: \forall t\}$   is independent of  $\{E_{i,t}: \forall t\}$ for all other $i\in \{1, \ldots, N\}/\{1\}$. 
\end{enumerate}

These conditions state that the  idiosyncratic noise of the initial period and of some individual  are  independent of the rest. Then we can use transformations of first observation and the observation of $i=1$ as the diversified weights: let 
$$
\bfb= (Y_{1,t_0},...,Y_{N,t_0})',\quad N\times 1; \quad 
\bfY_1= (Y_{1,1},...,Y_{1,T})',\quad T\times 1, \quad \text{and}
$$
$$
\bfW_{\bm\beta}= (\phi_1(\bfb),...,\phi_R(\bfb));\quad 
\bfW_\bfF= (\psi_1(\bfY_1),...,\psi_R(\bfY_1))
$$
where $\{\phi_k:k\leq R\}$ and $\{\psi_k:k\leq R\}$ are the sets of transformation functions. Then apply it to data except for $t=t_0$ and $i=1$. The cost would be the loss of $N+T$ observations in the panel, which is mild.



\section{Simulation results}\label{sec:simulation}

We evaluate the finite sample performance of our estimator in matrix completion design.  In  Section B of the supplement, we also examine the performance in other two designs:   varying coefficient model and heterogeneous treatment effect.  

The DGP for the matrix completion design is as follows: $$\bfY= \bfX \circ ({\bm\beta} \bfF' + \bfE^{\star} ) = \bfX \circ ({\bm\beta} \bfF')+\underbrace{\bfX\circ\bfE^{\star}}_{\coloneqq \bfE}$$
 where $\beta_{ik} =(2*\cos^k(b_i)+0.5*u_{ik})* N^{-(1-\alpha)/2}$, and $F_{tk}=2*\cos^k(f_t) +0.5*\nu_{tk}$ with $b_i$, $f_t$, $u_{ik}$, and $\nu_{tk}$ drawn from $\calN(0,1)$ independently. The noise $E^{\star}_{it}$ is generated from $\calN(0,1)$ independently. We note that the constant $\alpha \in (0,1]$ determines the strength of factors. A larger (smaller) $\alpha$ implies stronger (weaker) factors. $\bfX$ is the binary matrix where $X_{it} \sim \mathrm{Bernoulli}(p_i)$ with $p_i$ drawn from $\mathrm{Unif}[0.5, 0.8].$  Throughout simulations, we fix the true rank at $r=2$ and conduct $1,000$ replications.

We construct the diversified weights, $\bfW_{\bm\beta}$ and $\bfW_\bfF$ with $R=4$, following Section \ref{sec:choiceofW}, which are (1) Observed characteristics, (2)  Extra sample averages, and (3) Initial transformation. See Section B in the supplement for detailed implementation of each choice, the choice of the tuning parameter, and experiments with other values of $R$. We are interested in estimating three types of averages of the low-rank matrix: (I) ``Block," where $\calG=\calI \times \calT$ and $|\calI|=|\calT|=5$; (II) ``CS," where $\calG=\{1, \ldots, N\} \times \{T\}$, and (III) ``Serial," where $\calG=\{N\} \times \{1, \ldots, T\}$. 
 

\begin{table}[h]
\begin{center}
 \begin{tabular}{c|c|l|lll}
\toprule
\multicolumn{1}{l|}{Sample size} & \multicolumn{1}{l|}{Factor strength} & Diversified weights & Block  & CS     & Serial \\ \hline \hline
\multirow{6}{*}{$N=T=200$}         & \multirow{3}{*}{$\alpha=1$}        & Observed                & 0.932 & 0.940 & 0.950 \\
                                 &                                      & Extra                   & 0.913 & 0.940 & 0.946 \\
                                 &                                      & Initial                 & 0.874 & 0.945 & 0.951 \\ \cline{2-6} 
                                 & \multirow{3}{*}{$\alpha=0.5$}               & Observed         & 0.906 & 0.940 & 0.947 \\
                                 &                                      & Extra                   & 0.904 & 0.945 & 0.947 \\
                                 &                                      & Initial                 & 0.915 & 0.949 & 0.950 \\ \hline
\multirow{6}{*}{$N=T=400$}         & \multirow{3}{*}{$\alpha=1$}                 & Observed       & 0.942 & 0.954 & 0.952 \\
                                 &                                      & Extra                   & 0.921 & 0.943 & 0.948 \\
                                 &                                      & Initial                 & 0.884 & 0.945 & 0.944 \\ \cline{2-6} 
                                 & \multirow{3}{*}{$\alpha=0.5$}               & Observed         & 0.894 & 0.937 & 0.935 \\
                                 &                                      & Extra                   & 0.891 & 0.949 & 0.950 \\
                                 &                                      & Initial                 & 0.901 & 0.945 & 0.945 \\ \bottomrule
\end{tabular}
\end{center}
		\caption{\small Coverage probabilities of our debiased estimators in the matrix completion model. The target probability is 0.95. ``Observed,'' ``Extra,'' and ``Initial'' refer to our debiased estimators that use diversified weights of observed characteristics, the extra sample averages, and the initial transformation, respectively. Three averages are estimated: (I) ``Block," where $\calG=\calI \times \calT$ and $|\calI|=|\calT|=5$; (II) ``CS," where $\calG=\{1, \ldots, N\} \times \{T\}$, and (III) ``Serial," where $\calG=\{N\} \times \{1, \ldots, T\}$. In all specifications, we set $R=4$ while the true rank is $r=2$.}
  \label{table:MCtable}
\end{table}

Table \ref{table:MCtable} displays coverage probabilities of the confidence intervals, with the target probability 0.95. The weak factor setting ($\alpha=0.5$) is particularly relevant to our theory as the direct rank estimation   fails in this DGP.\footnote{The rank of the nuclear norm penalized estimator depends on the tuning parameter $\lambda.$ For the choice of $\lambda$, we follow the methods from \cite{chernozhukov2018inference,chernozhukov2023inference, choi2023inference}. The choice of $\lambda$ in these studies involves a choice of a small constant $c>0.$ \cite{chernozhukov2018inference,chernozhukov2023inference} set $c=1/10$ while \cite{choi2023inference} set $c=1/7.$   With both values, the rank of the nuclear norm penalized estimator is almost always one, whereas the  true rank $r$ is 2.} The coverage probabilities are reasonably good especially when estimating the ``CS" and ``Serial" averages, although we do observe size distortions  when estimating the ``Block" average.

\begin{figure}[h]
	\begin{center}
\includegraphics[width=0.9\textwidth]{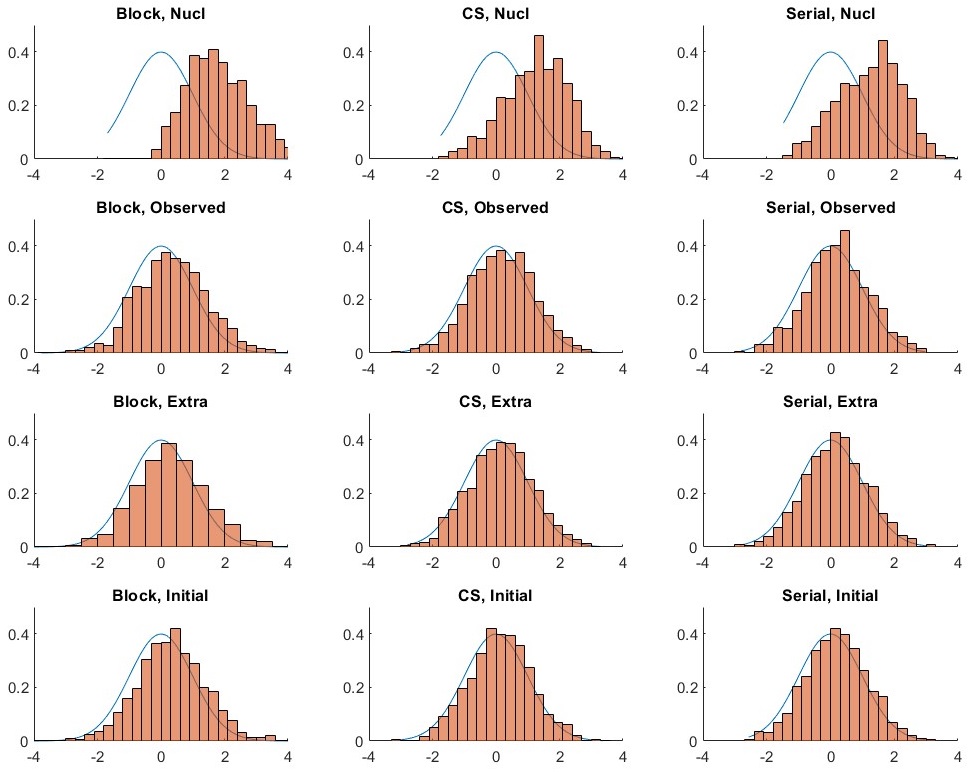} \vspace{-0.5cm}
	\end{center}
			\caption{\small Histograms of the standardized estimates in the noisy matrix completion model when $N=T=200$ and $\alpha=0.5$. ``Nucl'' in the first row presents the nuclear norm penalized estimators. The solid line represents the standard normal distribution. For our debiased estimators, we set $R=4$ while the true rank is $r=2$.}		\label{fig:MC200200alpha05}
		\centering
	\end{figure}

To compare among  choices of the weighting matrices,   we fix   $N=T=200$ and $\alpha=0.5$ (weak factors) and present the histograms of the standardized estimates, with the standard normal density, in Figure \ref{fig:MC200200alpha05}. For comparison, we include the  nuclear norm penalized estimator (``Nucl") whose  standard error is simulation-based. The standard error is estimated for all other methods. It is evident that ``Nucl" suffers from bias, and  our estimators approximate the standard normal distribution well for all group types and diversified weights.
 
In the supplement we  also compare the MSE of these methods. Not surprisingly,  ``Observed" and ``Extra" yield smaller MSE as they use extra information to specify the DP weights. But appealingly, the initial transformation, which does not require extra information, performs reasonably well.

\section{Empirical Study: Impact of the U.S. Presidential Election on the Federal Grants Allocation}\label{sec:empirical}


We apply the inferential theory presented in Section \ref{sec:treatmenteffect} to investigate the influence of the U.S. presidential election on the federal grant allocation to states. Although the specific allocation of federal funds is carried out by Congress, the U.S. president also wields notable influence in the grant allocation process. The role of the president in the allocation involves proposing annual federal budget proposals to Congress, signing or vetoing bills, and supervising the executive agencies. In the allocation process, the presidents may seek to allocate more federal funds to the states that supported them in the election, for political incentives. This practice is commonly referred to as ``pork-barrel politics.'' For decades, this practice has been investigated in numerous studies both theoretically and empirically \citep[e.g.,][]{larcinese2006allocating,berry2010president}. 
By employing the treatment effect estimation application presented in Section \ref{sec:treatmenteffect}, we aim to test whether the pork-barrel politics exist or not, in the history of the U.S. presidential elections.


We use the data of the U.S. federal grants, which cover fiscal years from 1953 to 2021 and include 50 U.S. states in addition to the District of Columbia.\footnote{In these data sets, the years 1960, 1972, and 1977 to 1980 are missing. As a result, our analysis does not include the Carter administration.} These data are publicly accessible on the websites of the U.S. Census Bureau, the National Associate of State Budget Officers (NASBO), and the Social Security Administration (SSA). 

Following the notation in Section \ref{sec:treatmenteffect}, we define state $i$ is ``treated" if it supported the incumbent president of year $t$ in the preceding election.  We note that the treatment assignments in this case could potentially be endogenous. Nevertheless, we presume that they are randomly assigned treatments and proceed to apply our approach. We calculate the per-capital federal grant, denoted as $Z_{it}$, for each state-year pair $(i,t)$. In order to detrend the data, we define $Y^{(\iota)}_{it}=X^{(\iota)}_{it} \times (Z_{it}/\sum_{i=1}^N Z_{it}) \times 100$ for each $\iota=0,1,$ and assume that $\bfY^{(\iota)}$ follows the model \eqref{eq:treatment}.

\begin{figure}[H]
	\begin{center}
\includegraphics[width= 0.8\textwidth, height=5cm]{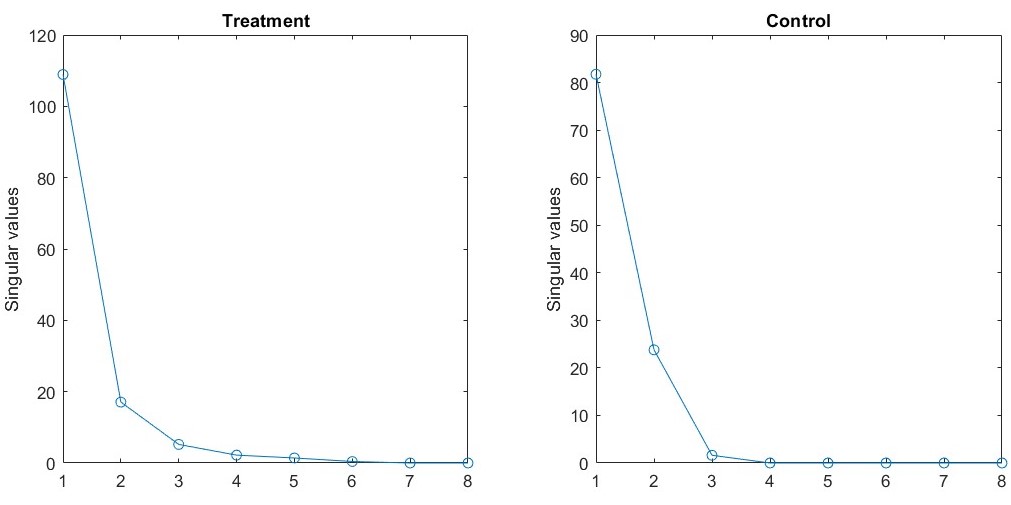} \vspace{-0.5cm}
	\end{center}
			\caption{\small The singular values of the (full-sample) nuclear norm penalized estimators for the treated and control sample, i.e., $\widetilde{\bfM}^{\full, (1)}$ and $\widetilde{\bfM}^{\full, (0)}$, respectively, in descending order. The left corresponds to the treated sample, while the right corresponds to the control sample.}	\label{fig:singularvalues}
	\end{figure}

Before proceeding to tests, we present the singular values of the nuclear norm penalized estimators for the treated and control samples (Figure \ref{fig:singularvalues}). The singular value plots do not provide a very definitive understanding of the true rank. While several data-driven methods choose $r=2$ as presented in Section A.3, we find that it does not lead to optimal out-of-sample performance in Section A.2. This motivates our method which does not rely on estimating the rank. 

We construct the diversified weighting matrices following the observed characteristics approach explained in Section \ref{subsec:obschr}. To be specific, the columns of $\bfW_\bfF$ and $\bfW_{\bm\beta}$ consist of polynomial transformations (up to the second power) of the annual data of U.S. GDP growth rates and unemployment rates,  and the state-by-state data on the averages of the population and the annual per-capita personal income from 1953 to 2021.

\begin{figure}[H]
	\begin{center}
\includegraphics[width= 1.0\textwidth, height=8.0cm]{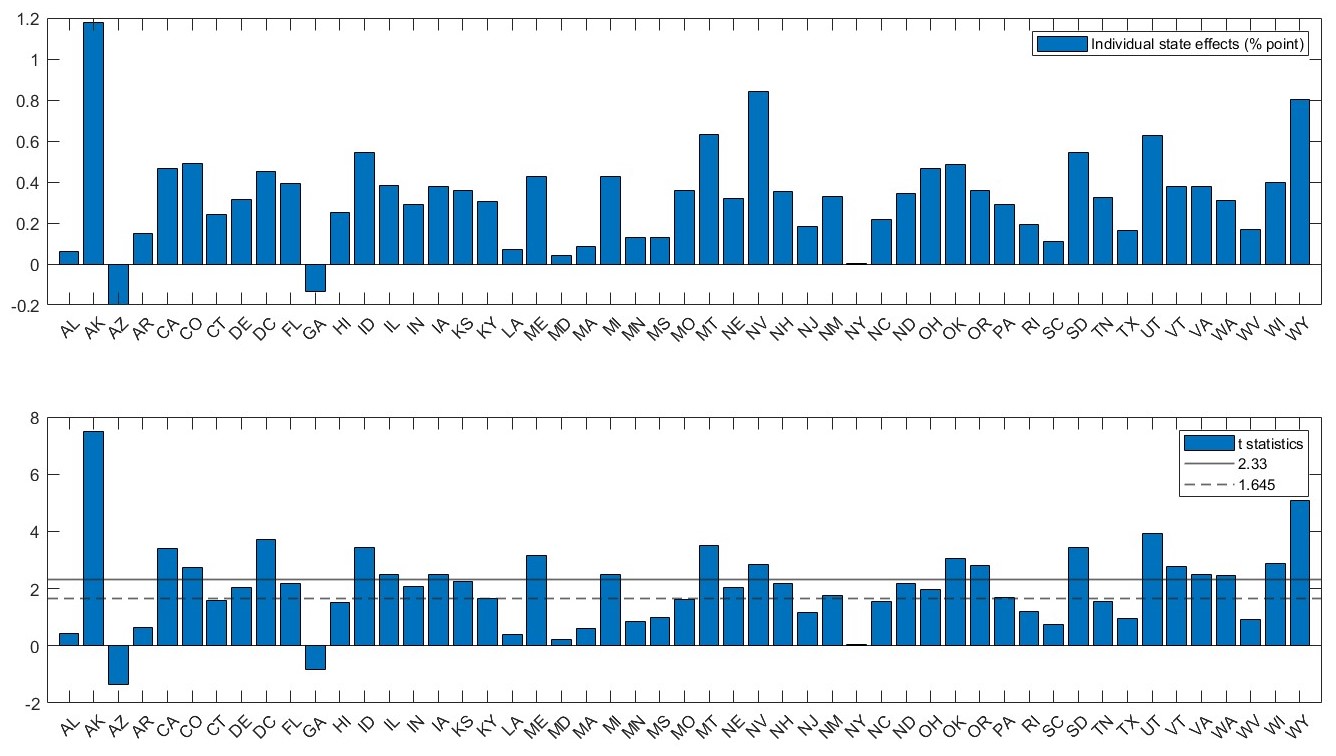} \vspace{-0.5cm}
	\end{center}
			\caption{Individual state effects and corresponding t-statistics}	\label{fig:individualstateeffects}
	\end{figure}

\begin{figure}[H]
	\begin{center}
\includegraphics[width=0.8\textwidth, height=7.7cm]{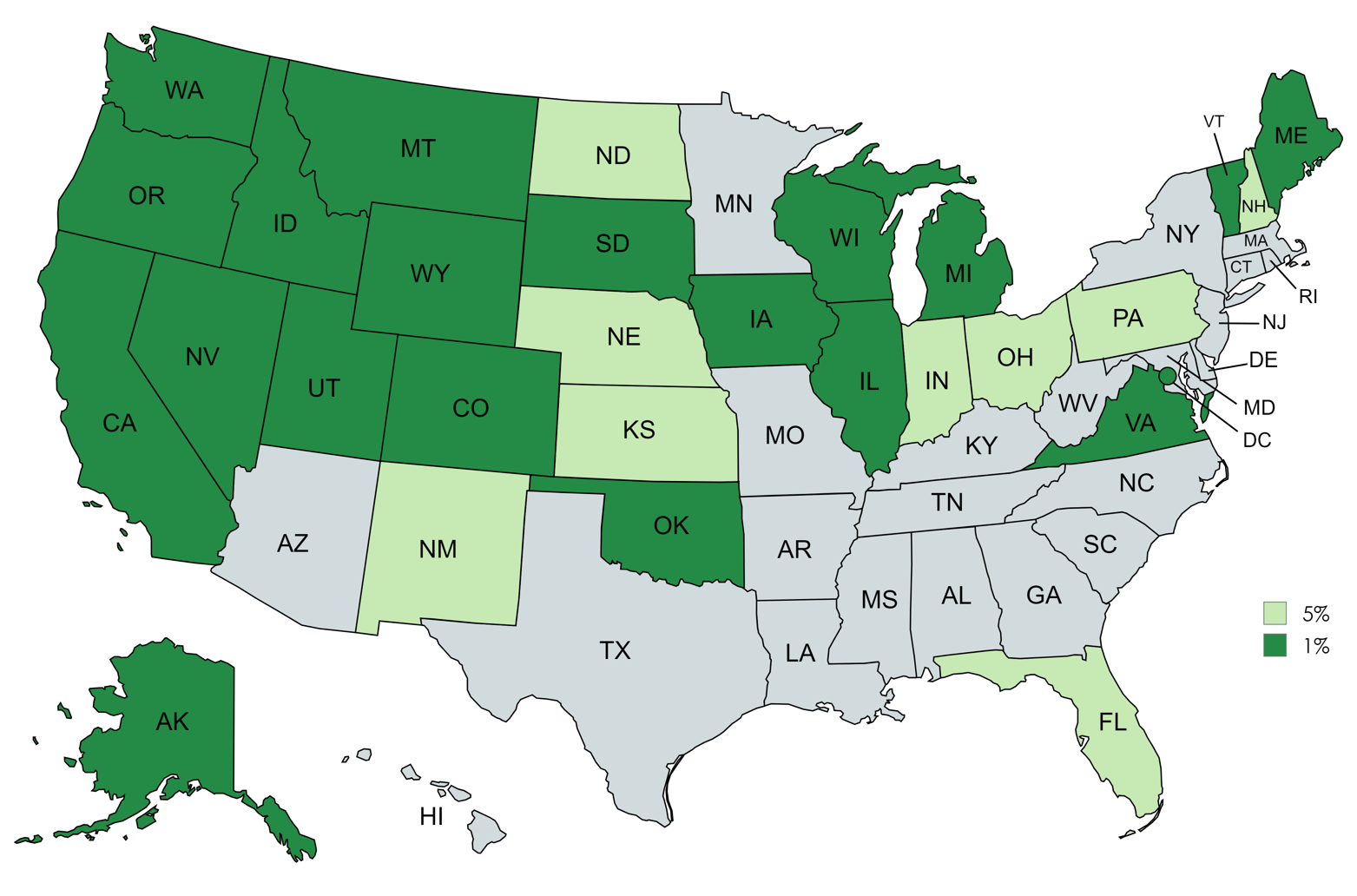} \vspace{-0.5cm}
	\end{center}
			\caption{\small The states in dark green indicate the rejection of the null hypothesis at a significance level of 1\%, whereas the states in light green correspond to a significance level of 5\%. In the gray-colored states, the null hypotheses are not rejected. This figure is created with \textsc{MapChart}.}	\label{fig:map}
	\end{figure}

To begin with, we estimate \textit{individual state effects}, i.e., the overall time average treatment effects for each state. The individual state effects and their   t-statistics are presented in Figure \ref{fig:individualstateeffects}. For each state, the null hypothesis is that there is no pork-barrel politics, i.e., the average treatment effect is non-positive.  We reject the null hypothesis in 30 states at the 5\% significance level and in 20 states at the 1\% significance level. Figure \ref{fig:map} depicts the distribution of the states where the null hypotheses are rejected. These test results suggest that the pork-barrel politics exist in a larger number of states. We now turn to a very natural question: Why are some states enjoying the ``pork,'' while others are not?

We observe that many of the states colored in dark green in Figure \ref{fig:map} are known as ``loyal'' states in that they have supported one party over decades. For example, DC has exclusively supported the Democratic Party since 1964, while AK, ID, OK, SD, UT, and WY have consistently supported the Republican Party since 1968. Inspired by this observation, we classify all 51 states based on their ``loyalty'' to a particular party. Specifically, we count the number of times that a state switches the party it supports, referred to as a ``swing,'' since the 1952 U.S. presidential election, in Table \ref{tab:numberofswing}.

\begin{figure}[H]
	\begin{center}
\includegraphics[width=1.0\textwidth, height=4.0cm]{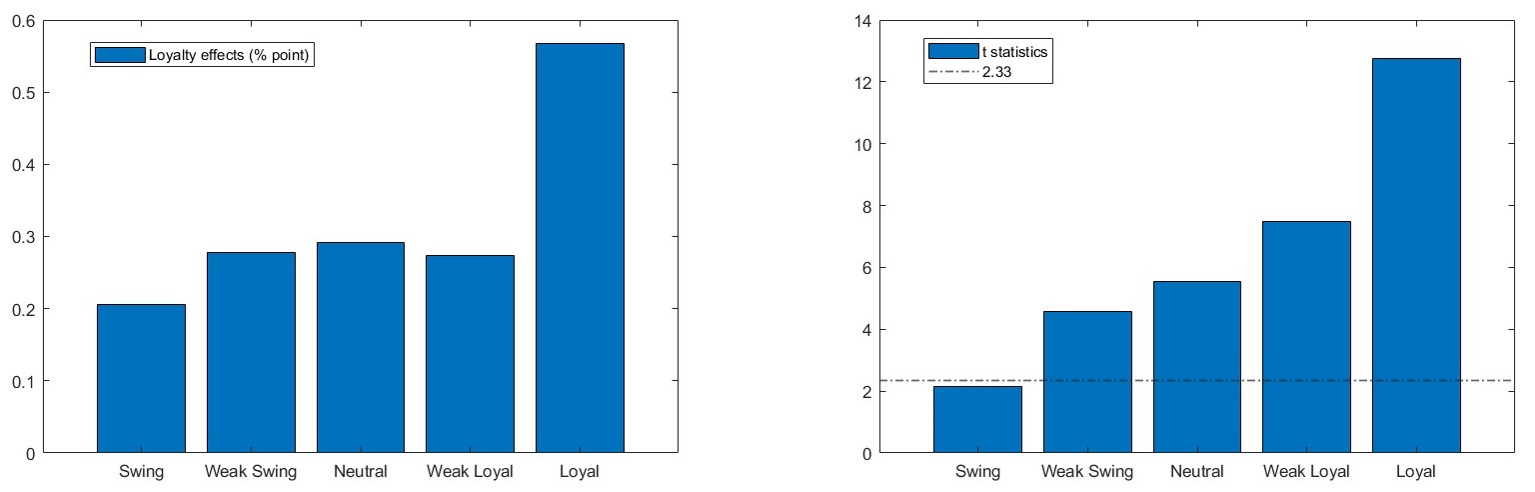} \vspace{-0.5cm}
	\end{center}
			\caption{\small Loyalty effects and corresponding t-statistics}	\label{fig:loyaleffects}
	\end{figure}

Based on Table \ref{tab:numberofswing}, we estimate the overall time average treatment effects for the state groups. The first plot in Figure \ref{fig:loyaleffects} indicates a positive relation between the loyalty and treatment effects: stronger loyalty leads to more substantial treatment effects. This ``rewarding-loyalty'' pattern is even clearer with t-statistics in the second plot. Also, these t-statistics  show that the pork-barrel politics exist in almost all groups. Only the t-statistic of the swing states is slightly lower than 2.33, and all other t-statistics are much larger. \footnote{In the appendix, we provide additional test results for other group averages: group averages for each Party governance and each presidential administration.}

 \begin{table}[H]    
  \begin{center} 
\begin{tabular}{lll}
\toprule
Group             & \# of swings & States                                                                                                      \\ \hline \hline
Swing states      & 8$\sim$      & FL, GA, LA, OH                                                                                              \\ \hline
Weak swing states & 6$\sim$7     & AR, IA, KY, MS, MO, PA, TN, WV, WI                                                                          \\ \hline
Neutral states    & 5            & AL, CO, DE, HI, MD, MI, NV, NH, NM, NY, NC, RI                                                              \\ \hline
Weak loyal states & 3$\sim$4     & \begin{tabular}[c]{@{}l@{}}AZ, CA, CT, IL, IN, ME, MA, MN,   MT, NJ, OR, \\ SC, TX, VT, VA, WA\end{tabular} \\ \hline
Loyal states      & 0$\sim$2     & AK, DC, ID, KS, NE, ND, OK, SD, UT, WY                                                                      \\ \bottomrule
\end{tabular}
\vspace{-0.5cm}
 	\end{center}
 	\caption{\small The counts of swings}\label{tab:numberofswing}
\end{table}

\onehalfspacing
 \small 
\bibliographystyle{apalike}
\bibliography{reference}

\end{document}